\documentstyle[12pt]{article}

\catcode`\@=11
\long\def\@makefntext#1{
\protect\noindent \hbox to 3.2pt {\hskip-.9pt
$^{{\ninerm\@thefnmark}}$\hfil}#1\hfill}		

\def\@makefnmark{\hbox to 0pt{$^{\@thefnmark}$\hss}}  

\def\ps@myheadings{\let\@mkboth\@gobbletwo
\def\@oddhead{\hbox{}
\rightmark\hfil\ninerm\thepage}
\def\@oddfoot{}\def\@evenhead{\ninerm\thepage\hfil
\leftmark\hbox{}}\def\@evenfoot{}
\def\sectionmark##1{}\def\subsectionmark##1{}}

\renewcommand{\thefootnote}{\fnsymbol{footnote}}

\newcounter{sectionc}\newcounter{subsectionc}\newcounter{subsubsectionc}
\renewcommand{\section}[1] {\vspace*{0.6cm}\addtocounter{sectionc}{1}
\setcounter{subsectionc}{0}\setcounter{subsubsectionc}{0}\noindent
	{\normalsize\bf\thesectionc. #1}\par\vspace*{0.4cm}}
\renewcommand{\subsection}[1] {\vspace*{0.6cm}\addtocounter{subsectionc}{1}
	\setcounter{subsubsectionc}{0}\noindent
	{\normalsize\it\thesectionc.\thesubsectionc. #1}\par\vspace*{0.4cm}}
\renewcommand{\subsubsection}[1]
{\vspace*{0.6cm}\addtocounter{subsubsectionc}{1}
	\noindent {\normalsize\rm\thesectionc.\thesubsectionc.\thesubsubsectionc.
	#1}\par\vspace*{0.4cm}}

\newcounter{appendixc}
\newcounter{subappendixc}[appendixc]
\newcounter{subsubappendixc}[subappendixc]

\renewcommand{\appendix}[1] {\vspace*{0.6cm}
        \refstepcounter{appendixc}
        \setcounter{figure}{0}
        \setcounter{table}{0}
        \setcounter{equation}{0}
        \renewcommand{\thefigure}{\Alph{appendixc}.\arabic{figure}}
        \renewcommand{\thetable}{\Alph{appendixc}.\arabic{table}}
        \renewcommand{\theappendixc}{\Alph{appendixc}}
        \renewcommand{\theequation}{\Alph{appendixc}.\arabic{equation}}
        \noindent{\bf Appendix \theappendixc #1}\par\vspace*{0.4cm}}

\def\abstracts#1{{
	\centering{\begin{minipage}{12.2truecm}\footnotesize\baselineskip=12pt\noindent
	\centerline{\footnotesize ABSTRACT}\vspace*{0.3cm}
	\parindent=0pt #1
	\end{minipage}}\par}}

\renewenvironment{thebibliography}[1]
	{\begin{list}{\arabic{enumi}.}
	{\usecounter{enumi}\setlength{\parsep}{0pt}
\setlength{\leftmargin 1.25cm}{\rightmargin 0pt}
	 \setlength{\itemsep}{0pt} \settowidth
	{\labelwidth}{#1.}\sloppy}}{\end{list}}

\newcounter{itemlistc}
\newcounter{romanlistc}
\newcounter{alphlistc}
\newcounter{arabiclistc}

\newcommand{\fcaption}[1]{
        \refstepcounter{figure}
        \setbox\@tempboxa = \hbox{\footnotesize Fig.~\thefigure. #1}
        \ifdim \wd\@tempboxa > 6in
           {\begin{center}
        \parbox{6in}{\footnotesize\baselineskip=12pt Fig.~\thefigure. #1}
            \end{center}}
        \else
             {\begin{center}
             {\footnotesize Fig.~\thefigure. #1}
              \end{center}}
        \fi}

\newcommand{\tcaption}[1]{
        \refstepcounter{table}
        \setbox\@tempboxa = \hbox{\footnotesize Table~\thetable. #1}
        \ifdim \wd\@tempboxa > 6in
           {\begin{center}
        \parbox{6in}{\footnotesize\baselineskip=12pt Table~\thetable. #1}
            \end{center}}
        \else
             {\begin{center}
             {\footnotesize Table~\thetable. #1}
              \end{center}}
        \fi}

\def\@citex[#1]#2{\if@filesw\immediate\write\@auxout
	{\string\citation{#2}}\fi
\def\@citea{}\@cite{\@for\@citeb:=#2\do
	{\@citea\def\@citea{,}\@ifundefined
	{b@\@citeb}{{\bf ?}\@warning
	{Citation `\@citeb' on page \thepage \space undefined}}
	{\csname b@\@citeb\endcsname}}}{#1}}

\newif\if@cghi
\def\cite{\@cghitrue\@ifnextchar [{\@tempswatrue
	\@citex}{\@tempswafalse\@citex[]}}
\def\citelow{\@cghifalse\@ifnextchar [{\@tempswatrue
	\@citex}{\@tempswafalse\@citex[]}}
\def\@cite#1#2{{$\null^{#1}$\if@tempswa\typeout
	{IJCGA warning: optional citation argument
	ignored: `#2'} \fi}}

 1
 1
 1

\font\ninerm=cmr9

\begin{document}

\centerline{\normalsize\bf Formation of Deeply Bound 1$s$ Pionic States
 in the $^{206}$Pb(d,$^3$He) Reaction}
\baselineskip=16pt

\vspace*{0.6cm}
\centerline{\footnotesize S. Hirenzaki$^{a)}$ and H. Toki$^{b)}$}
\baselineskip=16pt
\centerline{\footnotesize\it a) Department of Physics, 
Nara Women's University, Nara 630, Japan}
\baselineskip=12pt
\centerline{\footnotesize\it b) RCNP, Osaka University, Ibaraki, Osaka 567,
Japan}

\vspace*{0.9cm}
\abstracts{Recently, deeply bound pionic states were found experimentally 
in (d,$^3$He) reactions on $^{208}$Pb.  
The observed spectrum showed an excellent agreement with the DWIA 
calculation and the dominant peak was 
attributed to 
the pionic $2p$ state contribution.  
We studied theoretically 
$^{206}$Pb(d,$^3$He) reactions within the same model,  
and found that it is very likely to 
observe the pionic $1s$ state as an isolated 
peak in the $^{206}$Pb(d,$^3$He) reaction 
with feasible energy resolution and statistics.}

\vspace*{0.8cm}
\normalsize\baselineskip=15pt
\setcounter{footnote}{0}
\renewcommand{\thefootnote}{\alph{footnote}}

Since the suggestion of Toki and Yamazaki for the formation of deeply 
bound pionic atoms such as $1s$ and $2p$ states in heavy nuclei by direct 
reactions \cite{Toki88,Toki89},  there have been a number of efforts 
to find these states both experimentally and theoretically 
\cite{Iwasaki91}$^-$\cite{Matsuoka95}. 
It is 
worth mentioning that (n,d) reactions \cite{Yamazaki93} and (p,pp) 
reactions \cite{Matsuoka95} on the $^{208}$Pb target were able to 
identify some strength in the excitation function below the pion 
production threshold.  It was, however, not yet convincing 
to claim for pionic atom formation due to the 
lack of good resolution and statistical accuracy.  Very recently, 
Yamazaki $et.al.$ performed 
an experiment of (d,$^3$He) reactions on $^{208}$Pb with better 
resolution and statistical accuracy \cite{Yamazaki96}, and 
succeeded in identifing clearly a peak structure in the bound pion region.  We 
found also that the theoretical predictions \cite{Hirenzaki91} made before the 
experiment agree almost perfectly with the experiment.  
This agreement provides 
a strong confidence on the predictability of the theoretical model used.      

We have analyzed the latest data in more detail using our model \cite{Toki96} 
and found that the 
experimental peak structure consists of several contributions.  The 
largest contributions are due to $[l_\pi \otimes j_n^{-1}] = [ 2p \otimes 
p_{1/2}^{-1}]$ and $[ 2p \otimes p_{3/2}^{-1}]$.  Contributions from the 
deepest pionic $1s$ state, which is the most interesting, can be found 
only as a skewed shape of the experimental peak because the contributions 
from the pionic $1s$ state and 
$2p$ state could not be separated with the experimental energy 
resolution.  We expect to identify the pionic $1s$ contribution as a 
shoulder with much better energy resolution ($ FWHM \approx 200keV$) as 
shown in Fig. 1 (a).  
However, it is 
very difficult to realize such high resolution 
experimentally.  At higher incident energies, 
because of the matching condition of the momentum transfer, 
the $[ 1s \otimes i_{13/2}^{-1}]$ configuration makes an isolated peak.  
In this case, however, the excitation strength is small and the peak 
identification will be difficult \cite{Toki96}.  

In this paper we study the spectral shape and the 
excitation function of the $^{206}$Pb(d,$^3$He) 
reaction instead in order to observe the pionic $1s$ contribution as a peak 
with feasible experimental energy resolution.  Since the contribution of 
the $[ 2p \otimes p_{1/2}^{-1}]$ configuration in the case of $^{208}Pb$ 
makes an additional 
peak between the 
$[ 2p \otimes p_{3/2}^{-1}]$ and the $[ 1s \otimes p_{3/2}^{-1}]$
peaks, we may be 
able to remove this if we do not have 
contributions from the $p_{1/2}^{-1}$ neutron hole.  It is expected that the 
valence two neutron holes in the ground state of $^{206}$Pb are dominantly 
$p_{1/2}$. 
Hence, the assumption that the $p_{1/2}$ orbital in the ground state 
of $^{206}$Pb is empty may hold.  This is the reason why we 
consider $^{206}$Pb as a target nucleus to observe the pionic $1s$ 
contributions as an isolated peak.  

First we consider the $^{206}$Pb(d,$^3$He)$\pi^{- 205}$Pb reaction as the 
$^{208}$Pb(d,$^3$He)$\pi^{- 207}$Pb without contributions from 
two neutrons in the $p_{1/2}$ orbital for qualitative discussions.  
We will calculate 
realistic spectra later.  We have used the effective number approach for the 
theoretical calculation \cite{Toki91,Hirenzaki91,Hirenzaki96}.  In Fig.1 
(a), the calculated spectrum of $^{208}$Pb(d,$^3$He)$\pi^{- 207}$Pb is 
shown.  This calculated spectrum agrees with the data very well 
\cite{Yamazaki96,Toki96}.  We show the spectrum without the contribution from 
the $p_{1/2}$ neutron orbital in Fig.1 (b).  Clearly we can see the pionic 
$1s$ state contributions as an isolated peak at $Q \approx -134 MeV$ 
which consists of 
$[ 1s \otimes f_{5/2}^{-1}]$ and $[ 1s \otimes p_{3/2}^{-1}]$.  

We improve, then, the calculation by taking into account 
the realistic ground state configuration of $^{206}$Pb
, and the realistic excitation energies and strengths of $^{205}$Pb. 
The experimental data show that the 
$^{206}$Pb ground state is not a pure $p_{1/2}^{-2}$ state but 
contains admixtures of other two neutron hole configurations and can be 
written as \cite{Tickle69};

\begin{equation}
\Psi (^{206}Pb)_{g.s.} = a ( 2p_{1/2})^{-2} + b ( 1f_{5/2})^{-2}
+ c ( 2p_{3/2})^{-2} + d ( 0i_{13/2})^{-2}
\end{equation}

\noindent
with

\begin{displaymath}
a^2=0.54, b^2=0.20, c^2=0.12, d^2=0.12    .  
\end{displaymath}

\noindent
In our theoretical calculation, 
we assume the pion wavefunctions are the same as those of 
$\pi^{- 208}$Pb system and the neutron single particle wave 
functions are those of 
$^{208}$Pb.  
We calculate the effective numbers  
and normalize them using the occupation 
probability of each neutron state in the ground state of $^{206}$Pb.  
For example, there are $2$ $(= 2j + 1) $ neutrons in the $ 2p_{1/2} $ 
orbital in the ground state of $^{208}$Pb, while in the ground state of $ 
^{206}$Pb there are only $ 0.92$ $(= 2j + 1 - 2 a^2 )$ neutrons in 
average, where $a$ is the $( 2p_{1/2})^{-2}$ expansion coefficient in eq. 
(1).  Hence, for the $ 2p_{1/2} $ neutron orbital, the normalization factor is 
$0.46$ $( = 0.92/2 )$.  This normalization is necessary to get realistic 
total strength for all transitions corresponding to the pickup of a neutron in 
each orbital.  We show in Table 1 the normalization factor for each neutron state 
which appeared in eq. (1).  

As for the excited levels and excitation strengths of $^{205}$Pb, 
we use experimental excitation energies and spectroscopic factors 
obtained by $^{206}$Pb(d,t)$^{205}$Pb \cite{Tickle69}.  
Since one neutron 
pickup reaction from a certain orbital in $^{206}$Pb can be coupled 
to some excited 
states of  $^{205}$Pb, 
we need to distribute the effective numbers to each excited levels of $^{205}$Pb
in proportion to the experimental spectroscopic factors.  
In Table 2 we show the experimental 
excitation energies and spectroscopic factors.  We also show the relative 
strength of one neutron pickup processes from the same initial 
neutron state in the $^{206}$Pb, which should be multiplied to the 
effective numbers in order to include the effect of level splitting in the final 
$^{205}$Pb.  We should mention here that we treat the $ 2p_{1/2} $ and 
$ 2p_{3/2} $ neutron states together as one neutron state since 
the spin of the two 
$p$ wave excited states of $^{205}$Pb, whose excitation energies are 0.80 MeV and 
0.98 MeV, were not determined.  Thus, we normalized the total effective 
numbers of these two states together using the factor in Table 1 and, 
then, distribute them to five $p$ wave excited states of $^{205}$Pb in 
proportion to the experimental spectroscopic factors.  In Table 3 we show 
the calculated binding energies and widths of $\pi^{- 205}Pb$ atom using 
the same theoretical model as in ref. \cite{Toki89}.  

We will show an example using $[l_\pi \otimes j_n^{-1}] = [ 2p \otimes 
p_{1/2}^{-1}]$ and $[ 2p \otimes p_{3/2}^{-1}]$ configurations.  Their 
effective numbers at $T_d = 600 MeV$ were calculated to be $7.37 \times 
10^{-3}$ and $1.47 \times 10^{-2}$.  First they are normalized together 
for the ground state of $^{206}$Pb as $[7.37 \times 10^{-3} + 
1.47 \times 10^{-2}] \times 0.78 = 1.72 \times 10^{-2} $ using the 
normalization factor in Table 1.  This effective 
number is distributed to each excited state of $^{205}$Pb 
using the relative strength shown in Table 2.  For example, 
the effective number for a state with $E_x = 0.002 MeV$ is 
$1.72 \times 10^{-2} \times 0.234 = 4.02 \times 10^{-3}$.  We performed 
similar calculations for all other configurations.  

For pion quasi-free production and contributions from deeper neutron holes, 
we use 
the same spectrum as in Fig. 1 since it is not essential in this paper.  
Our model also implicitly postulates that the 
configuration mixing of the neutron-hole and pion-particle state in 
pionic atoms in $^{205}$Pb is small as in the case of pionic atoms in 
$^{207}$Pb. This assumption can be checked by knowing the width of the 
peak since the mixing must broaden the width.  In the latest data 
\cite{Yamazaki96}, the peak width is consistent with this assumption within 
the experimental energy resolution.  

We show the calculated results in Fig. 2 where negative pion production 
threshold with ground state of $^{205}Pb$ corresponds to $Q=-140.87MeV$.
We see that 
the pionic $1s$ contribution is also seen as an isolated peak in this 
realistic calculation.  The peak includes contributions from 
$(1s)_\pi (^{205}Pb)_{g.s.}$ and $(1s)_\pi (^{205}Pb)_{0.26MeV}$.  
These two 
contributions can't be distinguished in this case since the difference 
of the excitation energies of these two neutron levels (0.26MeV) 
are smaller than 
the width of the $1s$ pionic state due to the strong absorption effect. 
In Fig. 3 we show the same result as in 
Fig. 2 except for the 
different instrumental energy resolution.  Even with the feasible energy 
resolution, we can see the pionic $1s$ contributions as a peak as we 
expected in the beginning of this paper.  
Hence, this should be a good reaction to 
observe the deepest pionic state.  

We would like to comment here how we can choose the best incident energy 
for the pionic $1s$ state formation.  In our theoretical model the 
elementary cross section and the effective numbers provide the incident energy 
dependence of the $^3$He spectrum.  The elementary cross section begins 
to be finite  
from the pion production threshold and takes the maximum value around 
$T_d = 600 MeV$.  Then, it decreases with increasing energies.  The 
energy dependence of the effective numbers was studied systematically 
for (n,d) reactions \cite{Toki91} and it was 
found that the matching condition 
plays an important role.  In the $^{206}$Pb(d,$^3$He)$\pi^{- 205}$Pb 
reaction, the effective numbers for the pionic $1s$ state formation with 
$2p$ and $1f$ neutron hole states takes maximum value around $T_d = 600 
MeV$ where the elemantary cross section is maximum.  Thus, we think this 
incident energy for deuteron beam is best for the present case.  
For the (d,$^3$He) reaction with target nuclei which have valence neutrons 
in the $s$ state, the best energy will be smaller than $600MeV$ 
since the effective numbers for $[ 1s \otimes s^{-1}]$ configulation 
takes maximum value at zero momentum transfer.  

In summary we have calculated the excitation spectrum of 
the  $^{206}$Pb(d,$^3$He) reactions at $T_d = 600 MeV$ in order to 
observe the deepest pionic $1s$ 
state.  
We have 
shown that the contributions from the pionic $1s$ state can be seen as an 
isolated peak with the feasible experimental energy resolution.  The peak 
includes both contributions from $[ 1s \otimes 2p^{-1}]$ and 
$[ 1s \otimes 1f^{-1}_{5/2}]$.  Since the 
excitation energies of neutron hole states 
are known experimentally, we can get information on the pionic $1s$ state from 
this peak.    
We believe that this theoretical work motivates further experimental 
efforts to develop the physics of pionic atom spectroscopy.

We acknowledge the continuous encouragement and fruitful discussions with 
Prof. T. Yamazaki, who suggested us to look into $^{206}$Pb(d,$^3$He) 
reactions for observation of the $1s$ pionic state.  
We are grateful to Prof. R. S. Hayano for 
suggestions and discussions.  
We are also grateful to Prof. K. Kume for stimulating discussions.

\pagebreak
Table 1  

\noindent
The occupation probabilities of neutron single particle states in the 
ground state of $^{206}$Pb, which are used as the normalization factors 
of the effective numbers for (d,$^3$He) reactions.  The combined 
occupation probability of the $p$ state ($p_{1/2} + p_{3/2}$) is 0.78.  

\begin{center}
\begin{tabular}{cc} \hline
  $2p_{1/2}$ & 0.46 \\ 
  $2p_{3/2}$ & 0.94 \\
  $1f_{5/2}$ & 0.93 \\
  $0i_{13/2}$ & 0.98 \\ \hline
\end{tabular}
\end{center}

\vspace*{0.3cm}

Table 2

\noindent
Excitation energies and 
spectroscopic 
factors in $^{205}$Pb from $^{206}$Pb(d,t)$^{205}$Pb data \cite{Tickle69}.  
Relative 
excitation strengths of excited levels from the same 
initial neutron state in 
$^{206}$Pb are also shown.
Each level is assigned according to the TABLE V in ref. 13.  

\begin{center}
\begin{tabular}{cccc} \hline
Level & E$_x$ [MeV] & S & Relative Strength \\ \hline
  $p_{1/2}$ & 0.002 & 1.67 & 0.234 \\
  $p_{3/2}$ & 0.26 & 4.69 & 0.658 \\
  $p_{3/2}$ & 0.58 & 0.13 & 0.018 \\
  $p$ & 0.80 & 0.03 & 0.004 \\
  $p$ & 0.98 & 0.61 & 0.086 \\ 
      &      &      &       \\
  $f_{5/2}$ & 0.0 & 5.60 & 0.974 \\
  $f_{5/2}$ & 0.75 & 0.15 & 0.026 \\
      &      &      &       \\
  $f_{7/2}$ & 1.61 & 0.71 & 0.119 \\
  $f_{7/2}$ & 1.77 & 5.27 & 0.881 \\
      &      &      &       \\
  $i_{13/2}$ & 1.01 & 11.5 & 1.000 \\ \hline
\end{tabular}
\end{center}

\pagebreak

\vspace*{0.3cm}

Table 3

\noindent
Calculated binding energies and widths of $\pi^{- 205}Pb$ atom in unit of 
$keV$.

\begin{center}
\begin{tabular}{ccc} \hline
$nl$ & $E$ & $\Gamma$ \\ \hline
 1s & 7012 & 648 \\
 2s & 2977 & 188 \\
 2p & 5187 & 433 \\
 3s & 1648 & 81 \\
 3p & 2427 & 159 \\
 3d & 2857 & 92 \\
 4s & 1048 & 41 \\
 4p & 1413 & 75 \\
 4d & 1608 & 53 \\
 4f & 1575 & 1.0 \\
 5s & 727 & 24 \\
 5p & 926 & 41 \\
 5d & 1029 & 30 \\
 5f & 1010 & 0.9 \\
 6s & 536 & 15 \\
 6p & 655 & 24 \\
 6d & 716 & 18 \\
 6f & 704 & 0.6 \\ \hline
 \end{tabular}
\end{center}

\pagebreak
\noindent
Fig.1 

\noindent
(a) Calculated forward cross sections of 
the $^{208}$Pb(d,$^3$He) reactions 
at $T_d = 600 MeV$ with $200 keV$ experimental resolution.  
(b) Same figure as (a) without the contributions from $2p_{1/2}$ neutron state. 
In both figures, the vertical dashed line 
denotes the $\pi^-$ emission threshold energy and the flat background is 
assumed to be $20 [\mu b/sr/MeV]$.  

\vspace*{0.3cm}

\noindent
Fig.2

\noindent
Calculated forward cross sections of $^{206}$Pb(d,$^3$He) reactions at 
$T_d = 600 MeV$ with $200 keV$ experimental resolution, and 
with the use of the experimental neutron hole configurations and the 
excitation strengths.  Thin solid line 
shows the contributions from the neutron $2p$ states ( $p_{1/2}$ and 
$p_{3/2}$ ), and dashed curve 
those from the neutron $1f$ states ( $f_{5/2}$ and 
$f_{7/2}$ ).  

\vspace*{0.3cm}

\noindent
Fig.3

\noindent
Calculated forward cross sections of the $^{206}$Pb(d,$^3$He) reactions at 
$T_d = 600 MeV$ with $400 keV$ experimental resolution.

\end{document}